\documentclass{PoS}

\usepackage{subcaption}

\title{Unravelling the progenitors of merging black hole binaries}

\ShortTitle{Merging black hole binaries}

\author{\speaker{Nicola Giacobbo}\\
Dipartimento di Fisica e Astronomia ``G. Galilei'',
    Universit\`a di Padova. \\
        INAF, Osservatorio Astronomico di Padova, Padova, Italy. \\
        INFN, Milano Bicocca, Milano, Italy \\
        E-mail: \email{giacobbo.nicola@gmail.com}}
\author{Michela Mapelli\\
		Institut f\"ur  Astro- und Teilchenphysik, Universit\"at Innsbruck, Innsbruck, Austria \\
        INAF, Osservatorio Astronomico di Padova, Padova, Italy.\\
        INFN, Milano Bicocca, Milano, Italy \\
        E-mail: \email{michela.mapelli@oapd.inaf.it}}
\author{Mario Spera\\
		Institut f\"ur  Astro- und Teilchenphysik, Universit\"at Innsbruck, Innsbruck, Austria \\
        INFN, Milano Bicocca, Milano, Italy \\
        E-mail: \email{mario.spera@live.it}}

\abstract{The recent detection of gravitational waves has proven the existence of massive stellar black hole binaries (BBHs), but the formation channels of BBHs are still an open question. Here, we investigate the demography of BBHs by using our new population-synthesis code {\sc MOBSE}. {\sc MOBSE} is an updated version of the widely used binary population-synthesis code {\sc BSE}~\cite{Hurley2000,Hurley2002} and includes the key ingredients to determine the fate of massive stars: up-to-date stellar wind prescriptions and  supernova models. With {\sc MOBSE}, we form BBHs with total mass up to $\sim{}120$ M$_\odot$ at low metallicity, but only systems with total mass up to  $\sim{}80$ M$_\odot$ merge in less than a Hubble time. Our results show that only massive metal-poor stars ($Z\lesssim 0.002$) can be the progenitors of gravitational wave events like GW150914. Moreover, we predict that merging BBHs form much more efficiently from metal-poor than from metal-rich stars.}

\FullConference{GRAvitational-waves Science\&technology Symposium - GRASS2018\\
		1-2 March 2018\\
		Palazzo Moroni, Padova (Italy)}

\begin{document}

\section{Introduction}
The recent direct detection of gravitational waves (GWs,~\cite{Abbott2016d,Abbott2017a,Abbott2017b,Abbott2017c}) has revolutionized our knowledge about black holes (BHs). GW events confirm the existence of black hole binaries (BBHs) and prove that stellar mass BHs can be rather massive, with mass $\gtrsim 30$ M$_{\odot}$ as in the case of GW150914, GW170104 and GW170814. 
Despite the formation and evolution of BBHs have been studied for a long time, the state-of-the-art theoretical models still suffer from many uncertainties. This is particularly true for the final stages of massive star evolution. Some of the major issues are the treatment of  stellar winds and core-collapse supernovae (SNe). 
%The connection between the final stages of massive star evolution and the outcome of a SN explosion is a key point to understand BH formation, but the physics of core-collapse is extremely complex and still not fully understood. Even mass loss by stellar winds is crucial for the evolution of a massive star because it governs the final mass of a star. %Current models suggest that there is a strong relation between stellar winds and metallicity~\cite{Vink2001}. Moreover, it is apparent that mass loss also depends on the electron-scattering Eddington factor~\cite{Graefener2011}. If a star approaches the Eddington limit, the dependence of mass loss on metallicity tends to vanish. 
%Currently, there are only few population-synthesis codes containing up-to-date prescriptions for the evolution of massive stars. 
Here, we investigate the demography of BBHs by  using our new up-to-date population-synthesis code {\sc MOBSE} (which stands for "Massive Objects in Binary Stellar Evolution" \cite{Giacobbo2018}). 

\section{Description of {\sc MOBSE}}
{\sc MOBSE} is an updated version of the widely used binary population-synthesis code {\sc BSE}~\cite{Hurley2000,Hurley2002}. 
In {\sc MOBSE}, we introduced many upgrades with respect to {\sc BSE}, with the main purpose of improving the treatment of massive stars and stellar remnants. %Here we discuss our two major upgrades which mainly concern  the treatment of stellar winds and the prescriptions for SNe.

%We have implemented new recipes for stellar winds of hot massive stars in {\sc MOBSE}. %based on the most recent prescriptions of stellar evolution. In particular, %for O and B-type stars we use the formulas presented in~\cite{Vink2001}, for the Wolf-Rayet stars  we adopt the prescriptions described in~\cite{VinkdeKoter2005} and for the luminous blue variable stars we consider the recipes in~\cite{Belczynski2010}. 
%In general,

%%%%%%%%%%%%%%%%%%%%%%%%%%%%%%%%%%%%%FIGURE%%%%%%%%%%%%%%%%%%%%%%%%%%%%%%%%%%%
\begin{figure}
  \begin{subfigure}[b]{0.51\textwidth}
    \includegraphics[height=6.cm]{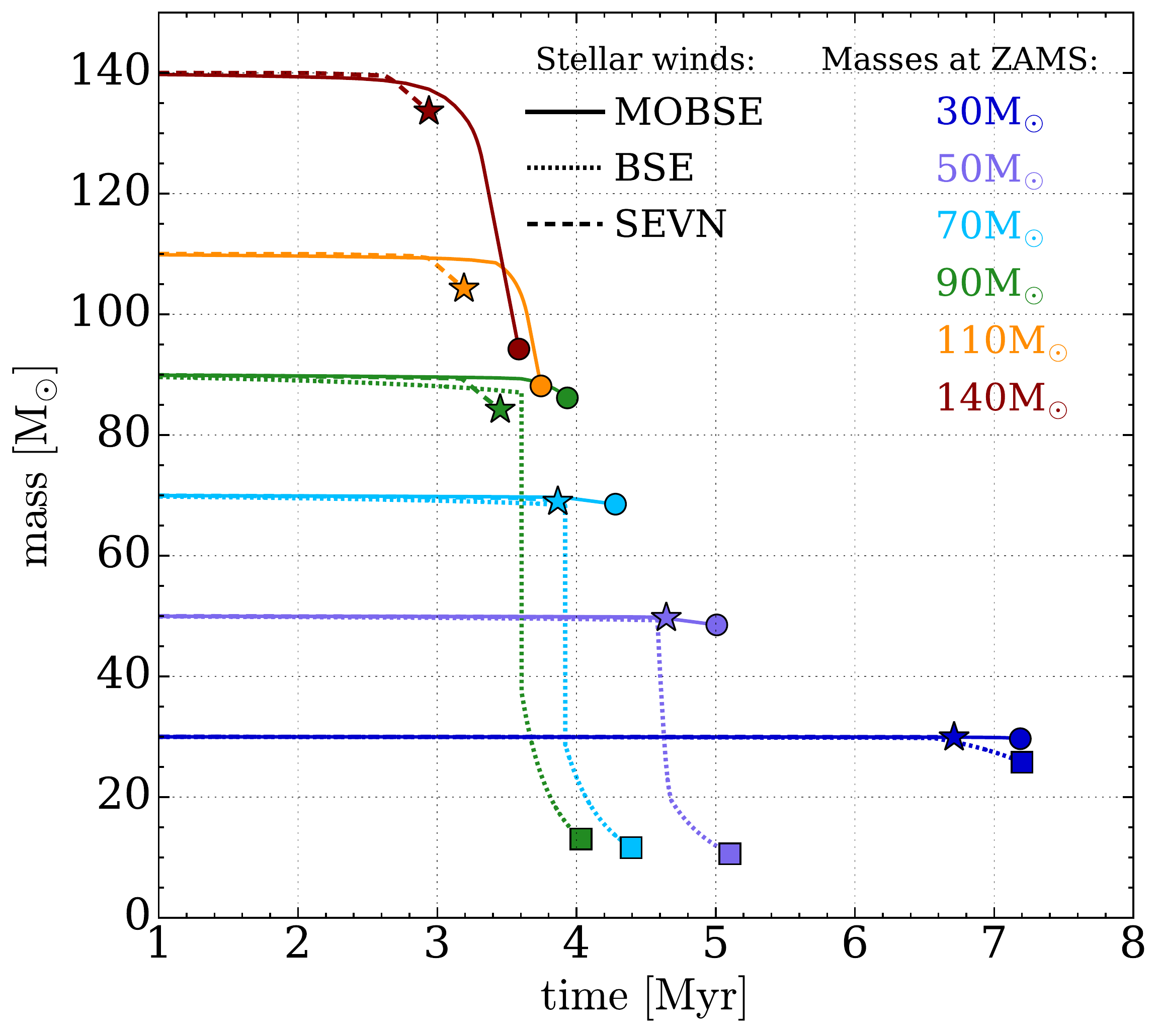}
  \end{subfigure}
  \begin{subfigure}[b]{0.51\textwidth}
    \includegraphics[height=6.cm]{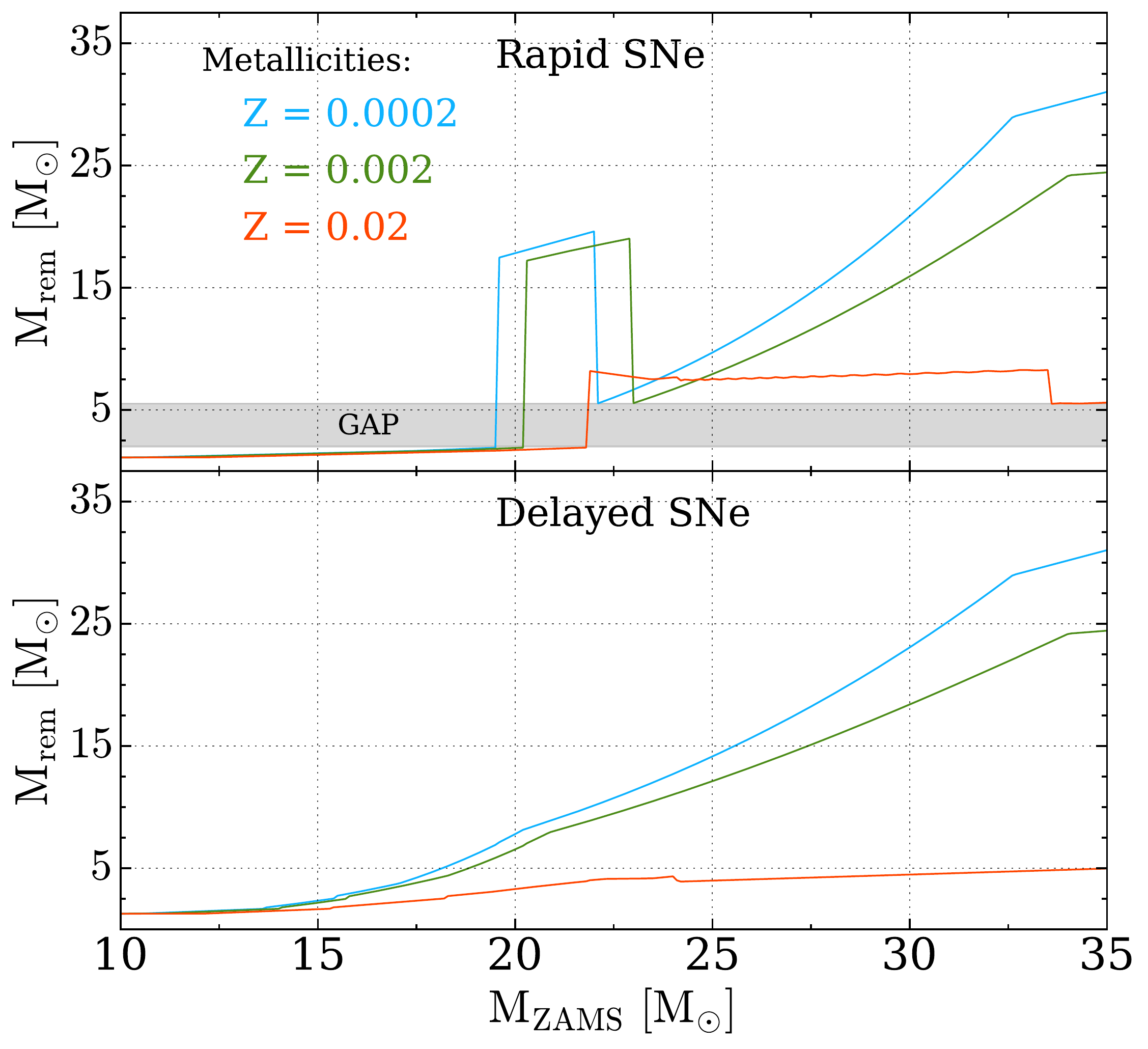}
    %\label{fig:SNe}
  \end{subfigure}
  \caption{\label{fig:winds}
a) Stellar mass as a function of time for six zero-age main sequence (ZAMS) masses at $Z = 0.0002$ computed with {\sc MOBSE} (solid lines), {\sc BSE} (dotted) and {\sc SEVN} (dashed). The markers identify the final mass of the stars: circles for {\sc MOBSE}, squares for {\sc BSE} and stars for {\sc SEVN}. b) Mass of the compact remnant ($M_{\rm rem}$) as a function of the ZAMS mass of the progenitor star ($M_{\rm ZAMS}$) for the rapid (upper panel) and delayed SN model (bottom panel). Red lines: $Z = 0.02$; green lines: $Z =0.002$; blue lines: $Z =0.0002$. In the top panel, the mass gap between the heaviest neutron stars and the lightest BHs ($\sim 2$ M$_{\odot}$ to $\sim 5$ M$_{\odot}$) is highlighted by a shaded area. ZAMS masses larger than $35$ M$_{\odot}$ are not shown because the outcomes of the rapid and delayed SN models are indistinguishable.}
\end{figure}
%%%%%%%%%%%%%%%%%%%%%%%%%%%%%%%%%%%%%%%%%%%%%%%%%%%%%%%%%%%%%%%%%%%%%%%%%%%%%%%
 
In particular, we model the mass loss of hot massive stars as $\dot{M} \propto Z^{\beta}$, where $Z$ is the metallicity.
The value of $\beta$ depends on the Eddington factor ($\Gamma_{\rm e}$): $\beta = 0.85$ if $\Gamma_{\rm e} < 2/3$, $\beta = 2.45 - 2.4\Gamma_{\rm e}$ if $2/3 \leq \Gamma_{\rm e} < 1$ and $\beta = 0.05$ if $\Gamma_{\rm e} \geq 1$ \cite{Chen2015}. Hence, when a massive star approaches and/or exceeds the Eddington limit its mass loss becomes almost insensitive to the metallicity. %Apart for {\sc SEVN}~\cite{Spera2015,Spera2017}, %The published version of SEVN can evolve only single stars but a new version that includes the main binary evolution processes is currently under develepment.} (acronym for "Stellar EVolution for N-body") {\sc MOBSE} is the only population-synthesis code that includes this dependence for the mass loss on the Eddington factor.

%The physics behind the core-collapse SNe is still uncertain and matter of debate. For that reason
We added in {\sc MOBSE} two prescriptions for the SN explosion, called {\it rapid} and {\it delayed} model (both described in details in~\cite{Fryer2012}). %In these models, the amount of fallback depends only on the final mass of the Carbon-Oxygen core and on the final mass of the star pre-explosion. %They are 
%Despite some studies~\cite{} suggest that there is a more complicated relation between the outcome of collapse and the properties of the star in {\sc MOBSE} the
 The main difference between these SN prescriptions is the time at which the explosion occurs: $t < 0.25$ s ($t \gtrsim 0.5$ s) for the rapid (delayed) model. %Since both these prescriptions do not distinguish between neutron stars (NSs) and BHs, we adopt the Tolman-Oppenheimer-Volko limit~\cite{Oppenheimer1939}: if a remnant is $\geq 3$ M$_{\odot}$ it is a BH otherwise a NS. 	
 Finally, {\sc MOBSE} includes recipes for pulsational pair-instability SNe  and for pair-instability SNe, as described in~\cite{Spera2017}. % and based on the results of~\cite{Woosley2017}.
  The reader can find more details about {\sc MOBSE} in~\cite{Mapelli2017,Giacobbo2018}.
 
%%%%%%%%%%%%%%%%%%%%%%%%%%%%%%FIGURE%%%%%%%%%%%%%%%%%%%%%%%%%%%%%%%%%%%%%%%%%
 \begin{figure*}
  \begin{center}
   \includegraphics[width=13cm]{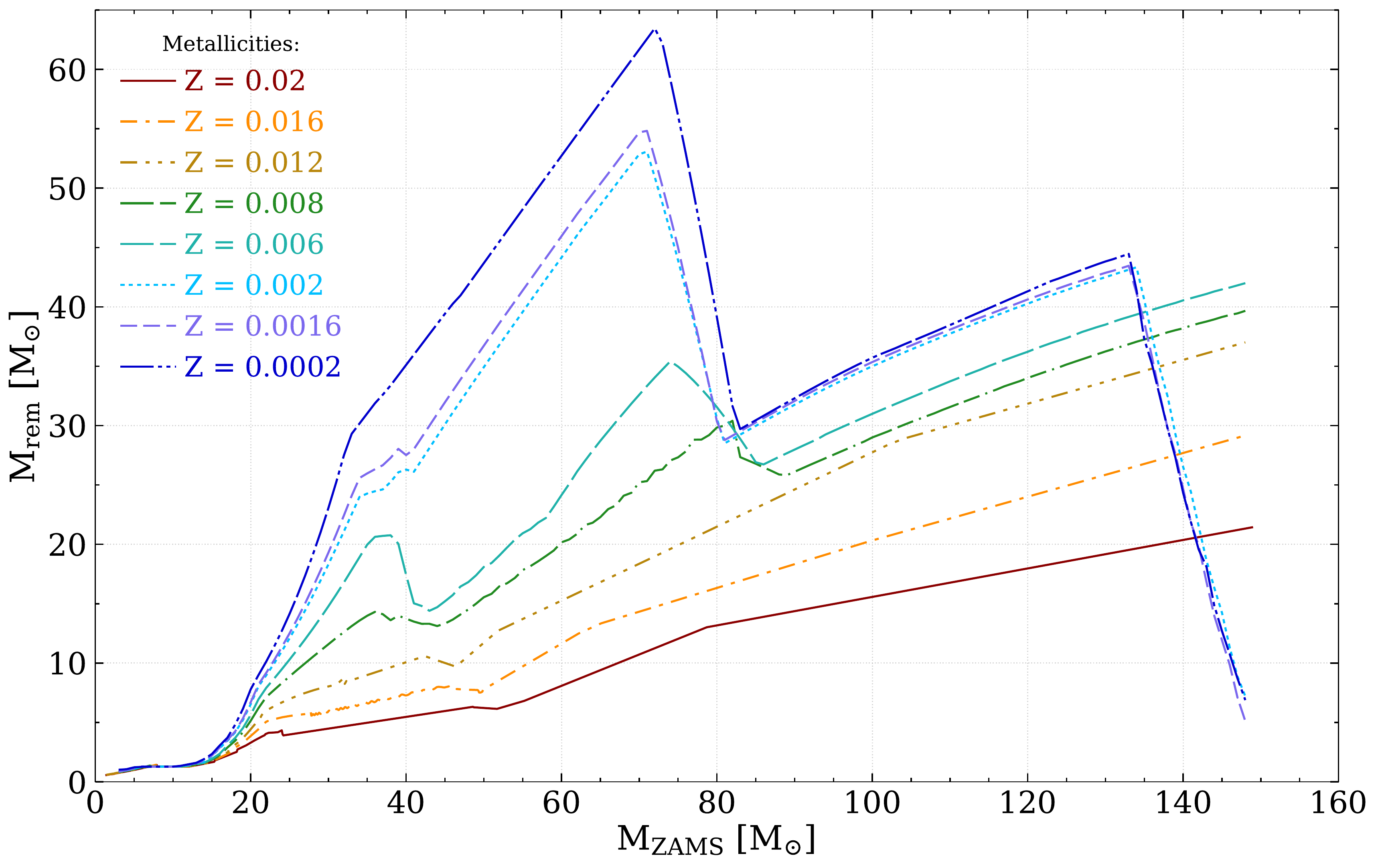}
   \caption{\label{fig:spec} Mass spectrum of the compact remnants ($M_{\rm rem}$) as a function of the zero-age main sequence (ZAMS) mass of progenitor stars, for different metallicity and adopting the delayed SN model.}
  \end{center}
 \end{figure*}

 %%%%%%%%%%%%%%%%%%%%%%%%%%%%%%%%%%%%%%%%%%%%%%%%%%%%%%%%%%%%%%%%%%%%%%%%%%%%%

\section{Results}
In the left-hand panel of Fig.~\ref{fig:winds} we compare the evolution of stellar mass at $Z=0.0002$ derived from {\sc MOBSE} with the evolution obtained with other two population-synthesis codes, namely {\sc BSE} and {\sc SEVN}~\cite{Spera2015,Spera2017}. 
%For low zero-age main sequence (ZAMS) masses ($\lesssim 30$ M$_{\odot}$) the evolution of stellar masses in different codes is almost indistinguishable.
For large zero-age main sequence (ZAMS) masses, the behavior of {\sc BSE}  is completely different with respect to both {\sc MOBSE} and {\sc SEVN}. This difference is due to stellar wind prescriptions. %Moreover, {\sc BSE} cannot evolve stars more massive than $100$ M$_{\odot}$.
The main difference between {\sc MOBSE} and {\sc SEVN} is the duration of stellar life. It is remarkable that {\sc MOBSE} and {\sc SEVN} behave so similarly, considering that {\sc MOBSE} computes single stellar evolution by using fitting formulas (described in \cite{Hurley2000}), while {\sc SEVN} uses up-to-date look-up tables based on {\sc PARSEC}~\cite{Chen2015}.
In the right-hand panels of Fig.~\ref{fig:winds}, we compare the rapid and delayed SN models. The main difference is that the rapid model predicts a remnant mass gap between $\sim 2$ M$_{\odot}$ and $\sim 5$ M$_{\odot}$.

Fig.~\ref{fig:spec} shows the mass spectrum of compact remnants as a function of the ZAMS mass for 8 metallicities (ranging from $Z = 0.0002$ to $Z = 0.02$), adopting the delayed SN model. It is apparent that there is a strong correlation between the maximum mass of the remnants and the metallicity. In particular, the lower the metallicity is, the higher the mass of the most massive remnant.
%PPISN and PISN

%%%%%%%%%%%%%%%%%%%%%%%%%%%%%%%%%%%FIGURE%%%%%%%%%%%%%%%%%%%%%%%%%%%%%%%%%%%%%%
\begin{figure*}
  \begin{center}
  \includegraphics[width=13cm]{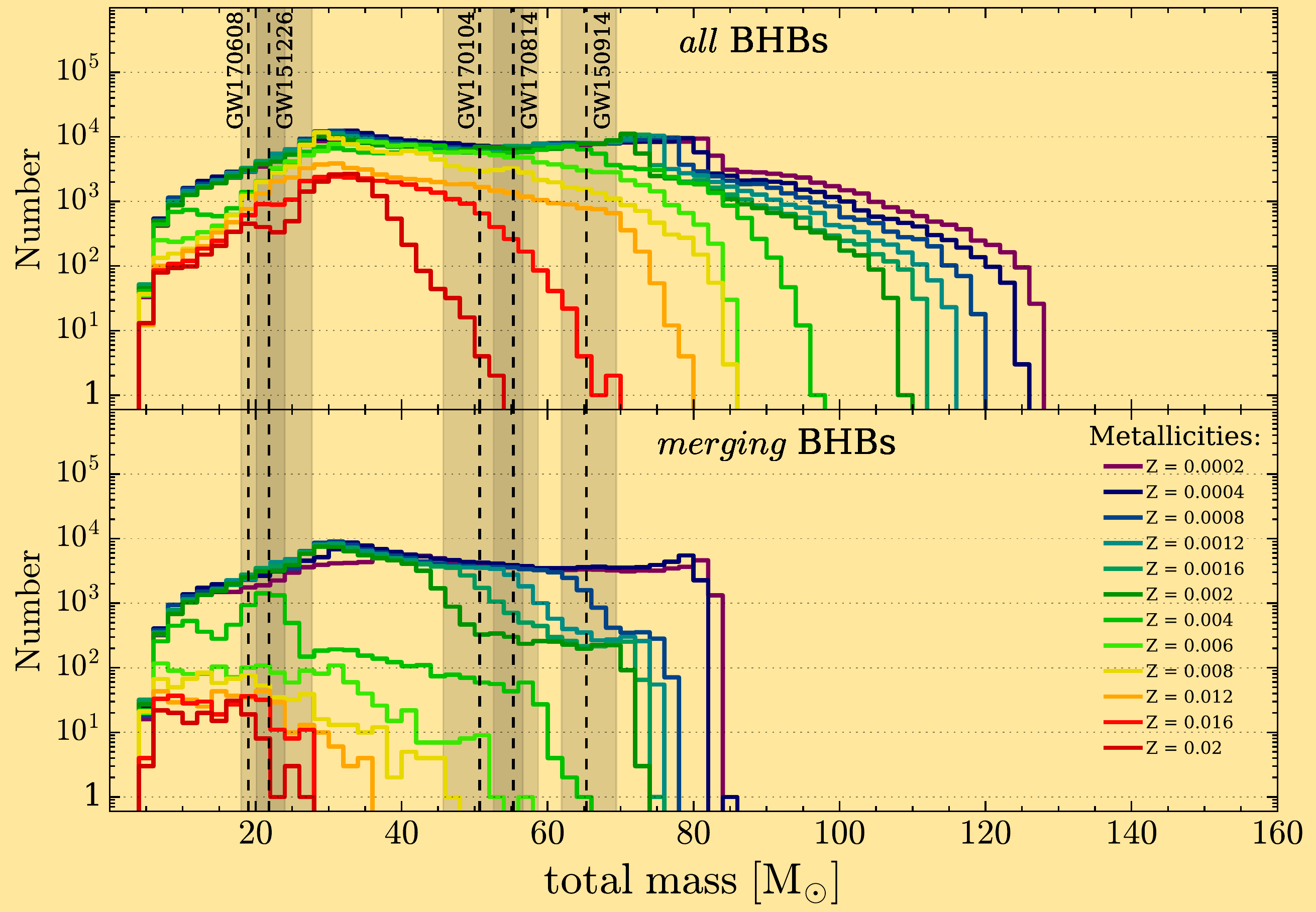}
  \caption{Total mass distribution for all BBHs (upper panel) and for the merging BBHs (lower panel) simulating $10^{7}$ binaries for each metallicity. Vertical dashed lines: total mass of GW150914, GW151226, GW170104, GW170608 and GW170814, with the corresponding 90 \%{} credible intervals (shadowed regions) \cite{Abbott2016d,Abbott2017a,Abbott2017b,Abbott2017c}.\label{fig:mtot}}
  \end{center}
  \end{figure*}

%%%%%%%%%%%%%%%%%%%%%%%%%%%%%%%%%%%%%%%%%%%%%%%%%%%%%%%%%%%%%%%%%%%%%%%%%%%%%%%

  %We use population-synthesis simulations performed with {\sc MOBSE} to study the demography of BBHs (see~\cite{Giacobbo2018,Mapelli2017} for more details).
  Fig.~\ref{fig:mtot} shows the distribution of total mass ($m_{1} + m_{2}$, where $m1$ and $m2$ are the masses of the primary and of the secondary BH, respectively) of BBHs formed in our simulations, considering different metallicities ($10^{7}$ binaries for each metallicity) and adopting the delayed SN model. In the top panel, we show the total mass distribution of all BBHs while in the bottom panel we show only the distribution for the sub-sample of systems merging within a Hubble time. As expected, the maximum mass of BBHs is higher at lower metallicity. From this Figure, it is also apparent that for a given metallicity the maximum mass of non-merging systems ($\sim 120$ M$_{\odot}$ at $Z=0.0002$) is significantly larger than the maximum mass of merging systems ($\sim 80$ M$_{\odot}$ at $Z=0.0002$). Moreover, the number of merging systems strongly depends on the metallicity, namely decreasing $Z$ the number of merging BBHs rises. 
Our results are consistent with the total mass of the five GWs detections of merging BBHs, but we find that only metal-poor stars can be the progenitors of the most massive merging systems like GW170104 ($Z \lesssim 0.006$), GW170814 ($Z \lesssim 0.004$) and GW150914 ($Z \lesssim 0.002$).
%We underline the fact that our results match perfectly with the total masses of all the first five gravitational detections merging BBHs.

\section{Summary}
We present our new population-synthesis code {\sc MOBSE} \cite{Mapelli2017,Giacobbo2018}. In {\sc MOBSE} we have implemented two prescriptions for SNe \cite{Fryer2012} and the most recent models of stellar winds \cite{Chen2015}. In particular, we included in {\sc MOBSE} also the effect of the Eddington factor on the mass loss. We used {\sc MOBSE} for studying the formation and evolution of BBHs. We find that the lower the metallicity is, the higher the maximum mass of BBHs. We form BBHs with total mass up to $\sim{}120$ M$_\odot$ at low metallicity, but only systems with total mass up to  $\sim{}80$ M$_\odot$ merge in less than a Hubble time.

The masses of our merging BBHs match those of the five reported BBH mergers. Only metal-poor progenitors can produce systems so massive as GW170104, GW170814 and GW150914. Finally, we find that merging BBHs form much more efficiently from metal-poor than from metal-rich progenitors.

\acknowledgments{NG acknowledges financial support from Fondazione Ing. Aldo Gini. MM  acknowledges financial support from the MERAC Foundation, from INAF through PRIN-SKA 'Opening a new era in compact object science with MeerKat', and from FWF through FWF stand-alone grant P31154-N27.} 
% This work benefited from support by the International Space Science Institute (ISSI), Bern, Switzerland,  through its International Team programme ref. no. 393 {\it The Evolution of Rich Stellar Populations \& BH Binaries} (2017-18).}


\begin{thebibliography}{99}

\bibitem{Abbott2016d} B. P. Abbott, R. Abbott, T. D. Abbott, M. R.Abernathy, F. Acernese, K. Ackley, C.  Adams, et al.
\emph{Binary Black Hole Mergers in the First Advanced LIGO Observing Run},
\emph{Physical Review X} {\bf 6} (4)
[{\tt arXiv:1606.04856}]

\bibitem{Abbott2017a} B. P. Abbott, R. Abbott, T. D. Abbott, F. Acernese, K. Ackley, C.  Adams, T. Adamas et al.
\emph{GW170104: Observation of a 50-Solar-Mass Binary Black Hole Coalescence at Redshift 0.2},
\emph{Physical Review Letters} {\bf 118} (22)
[{\tt arXiv:1706.01812}]

\bibitem{Abbott2017b} B. P. Abbott, R. Abbott, T. D. Abbott, F. Acernese, K. Ackley, C. Adams, P. Addesso et al.
\emph{GW170608: Observation of a 19 Solar-mass Binary Black Hole Coalescence},
\emph{ApJ} {\bf 851} (L35)
[{\tt arXiv:1711.05578}]

\bibitem{Abbott2017c} B. P. Abbott, R. Abbott, T. D. Abbott, F. Acernese, K. Ackley, C. Adams, T. Adams et al.
\emph{GW170814: A Three-Detector Observation of Gravitational Waves from a Binary Black Hole Coalescence},
\emph{Physical Review Letters} {\bf 119} (14)
[{\tt arXiv:1709.09660}]

%\bibitem{Vink2001} J. S. Vink, A. de Koter and H. J. G. L. M. Lamers,
%\emph{Mass-loss predictions for O and B stars as a function of metallicity},
%\emph{A \& A} {\bf 369} (574-588)
%[{\tt astro-ph/0101509}]

%\bibitem{Graefener2011} G. Gr{\"a}fener, J. S. Vink, A. de Koter, N. Langer,
%\emph{The Eddington factor as the key to understand the winds of the most massive stars. Evidence for a {$\Gamma$}-dependence of Wolf-Rayet type mass loss},
%\emph{A \& A} {\bf 535} (A56)
%[{\tt arXiv:1106.5361}]

\bibitem{Giacobbo2018} N. Giacobbo, M. Mapelli and M. Spera,
\emph{Merging black hole binaries: the effects of progenitor's metallicity, mass-loss rate and Eddington factor},
\emph{MNRAS} {\bf 474} (2959-2974)
[{\tt arXiv:1711.03556}]

\bibitem{Hurley2000} J. R. Hurley, O. R. Pols and C. A. Tout,
\emph{Comprehensive analytic formulae for stellar evolution as a function of mass and metallicity},
\emph{MNRAS} {\bf 315} (543-569)
[{\tt astro-ph/0001295}]

\bibitem{Hurley2002} J. R. Hurley, C. A. Tout and O. R. Pols,
\emph{Evolution of binary stars and the effect of tides on binary populations},
\emph{MNRAS} {\bf 329} (897-928)
[{\tt astro-ph/0201220}]





%\bibitem{VinkdeKoter2005} J. S. Vink and A. de Koter,
%\emph{On the metallicity dependence of Wolf-Rayet winds},
%\emph{A \& A} {\bf 442} (587-596)
%[{\tt astro-ph/0507352}]

%\bibitem{Belczynski2010} K. Belczynski and T. Bulik and C. L. Fryer and A. Ruiter and F. Valsecchi and J. S. Vink and J. R. Hurley,
%\emph{On the Maximum Mass of Stellar Black Holes},
%\emph{ApJ} {\bf 714} (1217)

\bibitem{Chen2015} Y. Chen, A. Bressan, L. Girardi, P. Marigo, X. 
	Kong, A. Lanza,
\emph{PARSEC evolutionary tracks of massive stars up to $350$ M$_{\odot}$ at metallicities $0.0001 \le Z \le 0.04$},
\emph{MNRAS} {\bf 452} (1068-1080)
[{\tt arXiv:1506.01681}]

\bibitem{Fryer2012} C. L. Fryer, K. Belczynski, G. Wiktorowicz, M. Dominik, V. Kalogera and D. E. Holz,
\emph{Compact Remnant Mass Function: Dependence on the Explosion Mechanism and Metallicity},
\emph{ApJ} {\bf 749} (91)
[{\tt arXiv:1110.1726}]


\bibitem{Spera2017} M. Spera and M. Mapelli,
\emph{Very massive stars, pair-instability supernovae and intermediate-mass black holes with the sevn code},
\emph{MNRAS} {\bf 470} (4739-4749)
[{\tt arXiv:1706.06109}]

%\bibitem{Woosley2017} S. E. Woosley,
%  \emph{Pulsational Pair-instability Supernovae},
%  \emph{ApJ} {\bf 836} (244-280)
%  [{\tt arXiv:1608.08939}]

\bibitem{Mapelli2017} M. Mapelli, N. Giacobbo, E. Ripamonti and M. Spera,
\emph{The cosmic merger rate of stellar black hole binaries from the Illustris simulation},
\emph{MNRAS} {\bf 472} (2422-2435)
[{\tt arXiv:1708.05722}]

\bibitem{Spera2015} M. Spera and M. Mapelli and A. Bressan,
\emph{he mass spectrum of compact remnants from the PARSEC stellar evolution tracks},
\emph{MNRAS} {\bf 451} (4086-4103)
[{\tt arXiv:1505.05201}]


%\bibitem{Bressan2012} A. Bressan, P. Marigo, L. Girardi, B. Salasnich, C. Dal Cero, S. Rubele and A. Nanni, \emph{New PARSEC evolutionary tracks of massive stars at low metallicity: testing canonical stellar evolution in nearby star-forming dwarf galaxies}, \emph{MNRAS} {\bf 445} (4287-4305) [{\tt arXiv:1410.1745}]

%\bibitem{Tang2014} J. Tang, A. Bressan, P. Rosenfield, A. Slemer, P. Marigo, L. Girardi, L. Bianchi, \emph{PARSEC: stellar tracks and isochrones with the PAdova and TRieste Stellar Evolution Code}, \emph{MNRAS} {\bf 427} (127-145) [{\tt arXiv:1208.4498}]






\end{thebibliography}
\end{document}